\documentclass[aps,twocolumn,superscriptaddress]{revtex4}

\usepackage{amsmath,amssymb,bm}
\usepackage{graphicx}
\usepackage{amsfonts}

\newcommand{\be}{\begin{equation}}
\newcommand{\ee}{\end{equation}}

\newcommand{\ba}{\begin{eqnarray}}
\newcommand{\ea}{\end{eqnarray}}

\begin{document}

%%%%%%%%%%%%%%%%%%%%%%%%%%%%%%%%%%%%%%%%%%%
%%%%%%%%%%%%%%%% the text %%%%%%%%%%%%%%%%%
%%%%%%%%%%%%%%%%%%%%%%%%%%%%%%%%%%%%%%%%%%%
\title{Anomalous pion production induced by
nontrivial  \\
topological structure of QCD vacuum}

\author{Nikolai Kochelev}
\email{kochelev@theor.jinr.ru} \affiliation{Institute of Modern
Physics, Chinese Academy of Science, Lanzhou 730000, China}
\affiliation{Bogoliubov Laboratory of Theoretical Physics, Joint
Institute for Nuclear Research,\\ Dubna, Moscow Region, 141980
Russia}

\author{Hee-Jung Lee} \email{hjl@chungbuk.ac.kr(correspondent)}
\affiliation{Department of Physics Education, Chungbuk National
University, Cheongju, Chungbuk 361-763, Korea}

\author{Baiyang Zhang}
\email{zhangbaiyang@impcas.ac.cn} \affiliation{Institute of Modern
Physics, Chinese Academy of Science, Lanzhou 730000, China}
\author{Pengming Zhang}
\email{zhpm@impcas.ac.cn} \affiliation{Institute of Modern Physics,
Chinese Academy of Science, Lanzhou 730000, China}

\begin{abstract}
A new mechanism for the pion
production in high energy reactions is suggested. It is related to
a possibility for the direct production of the pions induced by
instantons, topologically nontrivial gluonic excitations of the QCD
vacuum. This mechanism does not require
any fragmentation functions for the production of pseudoscalar mesons
in high energy reactions with hadrons.
We calculate the contribution of the new mechanism to the
inclusive $\pi^0$-meson production in high energy proton-proton
collisions. It is shown that it gives the dominant contribution
to the inclusive cross-sections in the  few GeV region for the
transverse momentum of the final pion with large rapidity. We
discuss the possible applications of the new mechanism to the
phenomenon of large spin effects observed in  numerous high energy
reactions and to
 particle productions in the relativistic heavy ion collisions.
\end{abstract}

\maketitle

\section{Introduction}
The inclusive production of mesons in high energy reactions is
one of the powerful tools to investigate the structure of
strong interaction. At the very large transverse momentum the
leading behavior of the inclusive cross-section should be
dominated by the perturbative t-channel one-gluon exchange which
leads to the well-known dependency $d\sigma\sim 1/p_t^4$. However,
one cannot believe in the validity of the perturbative QCD
(pQCD) approach in the small transverse momentum region where the
nonperturbative QCD effects can play a crucial role. It is clear
that the value of the transfer momentum for the applicability of the
pQCD depends on the nonperturbative dynamics of QCD. This dynamics
is deeply related to the complicated structure of the QCD vacuum.
One of the powerful models to calculate the nonperturbative QCD
effects in hadron physics, which are induced  by the nontrivial
topological structure of the QCD vacuum, is  the  instanton model (see
reviews \cite{shuryak,diakonov}). The instantons describe the
subbarrier transitions between the classical QCD vacua with the
different topological charge. The existence of instantons is very
important for hadron physics. For example, they provide a natural
mechanism of the spontaneous chiral symmetry breaking (SCSB) in the
strong interaction.  As the result, large dynamical quark masses
arise. One of the places where the SCSB effects might be
important is the high energy reactions with hadrons. In
particular, it was demonstrated in \cite{Kochelev:2006ny} that in
the few GeV region for the momentum transfer the instanton
effects give a significant contribution to the high energy
quark-quark scattering cross section. These effects come from the
anomalous chromomagnetic quark-gluon interaction induced by
instantons \cite{kochelev1}.
Furthermore, this interaction leads to the quark chirality flip
and might give an important contribution to the spin-dependent
cross sections \cite{kochelev2,kochelev3,shuryak2,zahed}. The
problem with the pQCD description of the high energy inclusive
pion production was discussed for the first time in \cite{soffer}.
It was mentioned that the cross section is well described by the pQCD
only in the region of small $x_F$ for the fixed target
experiments. In \cite{alesio}, the large partonic intrinsic
momentum in hadrons was introduced to describe the data. However,
it is not easy to justify the existence of such a large intrinsic
partonic momentum from the confinement dynamics. The possible
violation of the pQCD factorization in inclusive production of
hadrons induced by high twist contributions was discussed by
Brodsky with collaborators (see \cite{Arleo:2009ch,brodsky} and
references therein), but the microscopic mechanism of such
a violation was not presented.
\\
In this Letter, we suggest a new mechanism of the inclusive
production of pions in high energy reactions. We call it the
anomalous pion production (APP) because the formation of the pions
happens at short distances due to instantons
which have a much smaller size in comparison with the confinement
scale. Furthermore, in our approach it is not needed to include in
the calculation any fragmentation functions which are related to the
hadronization and, therefore, to the confinement dynamics. As the
result, the value of the inclusive pion cross section is
determined by the structure of the short range fluctuations of
gluonic fields in the QCD vacuum. This mechanism breaks  the pQCD
factorization and might be a corner stone of various phenomena
observed in high energy reactions in the few GeV range for
the transfer momentum.

\section{Anomalous production of the pions  induced by chromomagnetic vertex}
It was shown that instantons generate a new type of the quark-gluon
chromomagnetic interaction \cite{kochelev3}
\begin{equation}
{\cal L}_I=
i\frac{g_s\mu_a}{4M_q}\bar q\sigma^{\mu\nu}t^a q G^{a}_{\mu\nu},
\label{Lag1}
\end{equation}
where  $\mu_a$ is the anomalous quark chromomagnetic moment
(AQCM), $M_q$ is the dynamical quark mass, $g_s$ is the strong
coupling constant, and $G^{a}_{\mu\nu}$ is the gluon field
strength. Within the instanton model the
value of AQCM is (see for details \cite{diakonov,kochelev4})
\be
\mu_a=-\frac{3\pi (M_q\rho_c)^2}{4\alpha_s(\rho_c)}.
\label{AQCM1}
\ee
Therefore, the  value of AQCM  is determined by
the dynamical mass of the quark in the instanton vacuum. For
example, for the dynamical quark mass  $M_q=170$ MeV from the mean
field approximation \cite{shuryak} and
$\alpha_s(\rho_c) \approx 0.5$ at an average
size of instantons in the QCD vacuum $\rho_c=1/600 $ MeV$^{-1}$ \cite{diakonov},
we obtain
\begin{equation}
{\mu_a}=-0.378,
\label{mu}
\end{equation}
which is very large in comparison with  the Schwinger-type of the
pQCD contribution to the AQCM
\begin{equation}
\mu_a^{pQCD}=-\frac{\alpha_s}{12\pi}\approx -1.3\times 10^{-2}.
\end{equation}
It is evident that Lagrangian, Eq.\ref{Lag1}, violates chiral
symmetry. Therefore,  in  \cite{diakonov,Balla:1997hf} the
generalization of  Eq.\ref{Lag1} was suggested by inclusion of the pion field into
consideration to preserve the explicit chiral invariance.
The modified Lagrangian is
\begin{equation}
{\cal L}_I= i\frac{g_s\mu_a}{4M_q}\bar q\sigma^{\mu\nu}t^a
e^{i\gamma_5\vec{\tau}\cdot\vec{\phi}_\pi/F_\pi}q G^{a}_{\mu\nu},
\label{Lag2}
\end{equation}
where $F_\pi=93$ MeV.
\begin{figure}[htbp]
\centering{\includegraphics[width=7cm,height=2cm,angle=0]{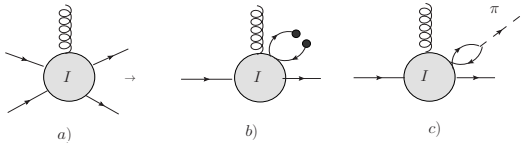}}
\caption{The diagram a) presents the general quark-gluon vertex generated by instantons for $N_f=2$ case ,
b) corresponds to the case with one of the quark lines connected through the
quark condensate, and c) describes the direct production of the pion from the instanton.}
\label{fig:1}
\end{figure}
The expansion of this Lagrangian up to the first order in the pion
field gives \footnote{We will neglect high order terms.  Their
contribution to the cross section is expected to be suppressed in
a large $N_c$ limit by a factor of  $1/N_c$, because $F_\pi\sim
\sqrt{N_c}$. Additionally, due to increasing of the number of the final particles,
it should be suppressed at large $x_F$ in comparison with a leading contribution.}
%Furthermore, in the single instanton approximation, which we are
%using, only two terms of the Eq.6 can contribute. High order pion
%field contributions coming from the multiinstanton configurations
%is expected to be suppressed by the small instanton packing
%fraction in the QCD vacuum $f\approx 0.1$ \cite{shuryak}. For the
%calculation of the high order contributions, one needs also to
%include $\sigma$ meson into the consideration.
%This task is beyond the scope of our paper.}
\begin{equation}
{\cal L}_I=
i\frac{g_s\mu_a}{4M_q}\bar q\sigma^{\mu\nu}t^a q G^{a}_{\mu\nu}-
\frac{g_s\mu_a}{4M_qF_\pi}\bar q\sigma^{\mu\nu}t^a \gamma_5\vec{\tau}\cdot\vec{\phi}_\pi q G^{a}_{\mu\nu}.
\label{Lag3}
\end{equation}
Within the instanton model  the two terms in this equation can be
represented by the diagrams b), c) in Fig.1. The diagrams which give
the contribution to the anomalous production of the  pion on the
parton level are shown in Fig.2.
\begin{figure}[htbp]
\centering{\includegraphics[width=7cm,height=3cm,angle=0]{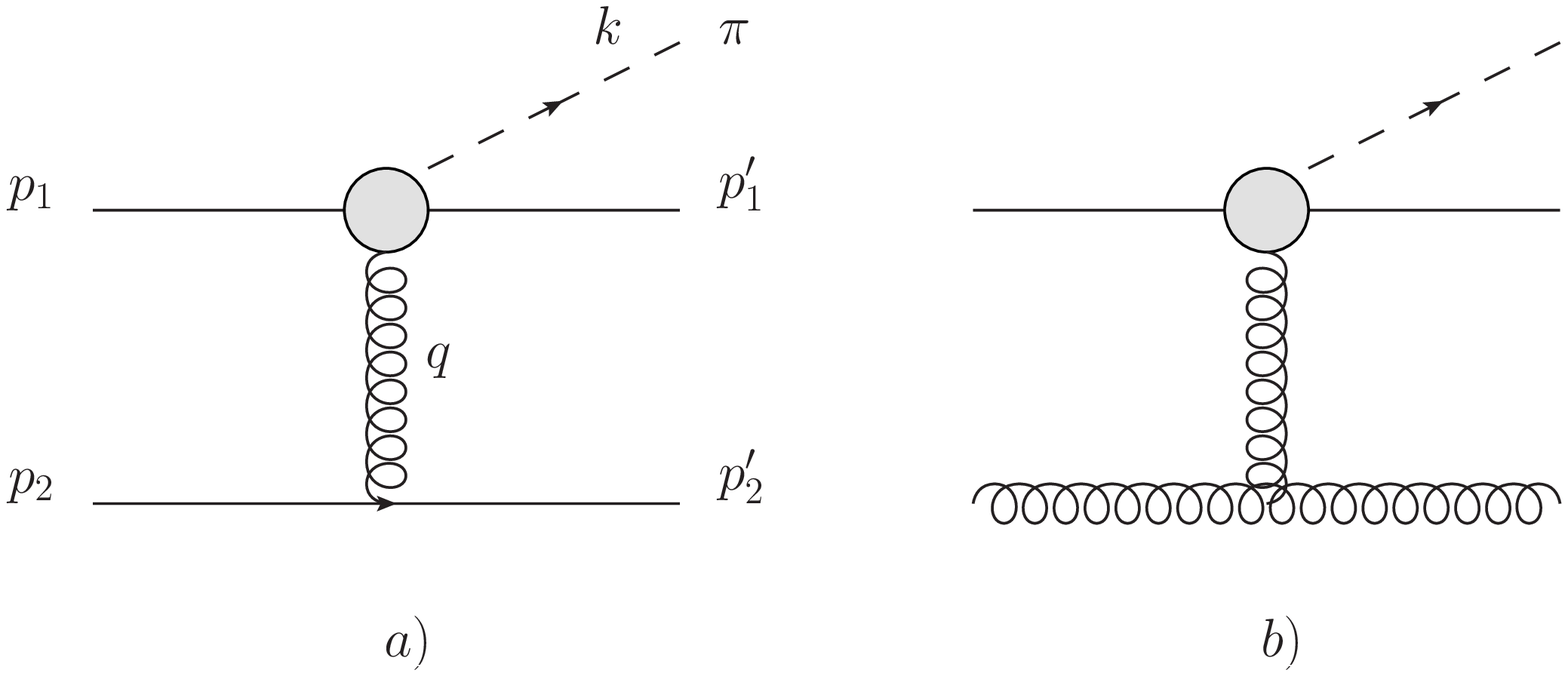}}
\caption{The pion production induced by instanton a) in quark-quark scattering
and b)in quark-gluon scattering.}
\end{figure}
Using the Sudakov parametrization of the  four-momenta of
particles \cite{Baier:1980kx}
\begin{equation}
p_i^\prime=\alpha_ip_2+z_ip_1+p_{i,t}^\prime, \ \  p_{i,t}^\prime\cdot p_{1,2}=0, \ \  p_{i,t}^{\prime2}=-\vec{p}_i^{\prime2}<0,
\nonumber
\end{equation}
we obtain the cross section of the  $\pi^0$ production in the
quark-quark scattering by incoming $u$- or $d$-quark \footnote{The
production cross section of  $\pi^+$ ($\pi^-$) in forward  $u$-
($d$-) quark scattering is twice larger.}
\begin{eqnarray}
&&\frac{zd\sigma^{qq \rightarrow qq\pi^0}}{dzd^2\vec{k}}
\nonumber\\
&=&g_{\pi^0qq}^2\frac{\rho_c^4}{2^7\pi}\int\frac{\alpha_s^2
(q^2)}{\alpha_s^2(\rho_c)}
\frac{{\vec q}^{2}}{({\vec q}^{2}+m_g^2)^2}F^2_g(\rho_c|\vec{q}|)d^2\vec{q},
\end{eqnarray}
where $F_g(t)=4/t^2-2K_2(t)$ is the form factor of the instanton
\cite{kochelev4}, $z$ is a fraction of the  initial quark momentum
carried by the final pion, $m_g$ is the dynamical gluon mass
related to the infrared behavior of the gluon propagator, and
 \be
g_{\pi^0qq}=\frac{M_q}{F_\pi}.
\label{GT}
\ee
For the quark-gluon scattering (the diagram in Fig.2b), we obtained the cross section
which is larger than quark-quark scattering case by a factor of
$9/4$. The instanton corresponds to the
subbarrier transition between vacua with the different topological
charge. Therefore, the single instanton approximation, which we are using,
is correct when the invariant mass of the partonic system produced by
the instanton does not exceed the height of the
potential barrier between these vacua. This height is given by the
energy of  the so-called sphaleron
$E_{sph}=3\pi/(4\alpha_s(\rho)\rho)$ (see, for example,
\cite{diakonov}). For $1/\rho_c=0.6$ GeV and $\alpha(\rho_c)=0.5$
we obtain  $E_{sph}=2.83$ GeV, which is rather a large value. By
using the condition $ (k+p_1^\prime)^2=M_X^2\leq E_{sph}^2$ and
the relation \be (z\vec{q}-\vec{k})^2=z(1-z)M_X^2, \label{rel}
\end{equation}
where $\vec{k}$ is the transverse momentum of the pion, we finally obtain
the APP cross section in the quark-quark scattering
\begin{widetext}
\begin{eqnarray}
\frac{zd\sigma^{qq \rightarrow qq\pi^0}}{dzd^2\vec{k}} =
g_{\pi^0qq}^2\frac{\rho_c^4}{\alpha_s^2(\rho_c)2^8\pi}
\int_0^{y_{max}}dy\int_0^{2\pi}d\varphi \frac{\alpha_s^2(x,y,\varphi)F(x,y,z,\varphi)
F^2_g(\sqrt{F(x,y,z,\varphi)}/z)}{(F(x,y,z,\varphi)+z^2m_g^2\rho_c^2)^2}z(1-z),
\label{partonicAPP}
\end{eqnarray}
\end{widetext}
where
\be
F(x,y,z,\varphi)=z(1-z)y+x^2+2x\sqrt{y}\sqrt{z(1-z)}\cos\varphi
\label{fun}
\ee
and $x=\rho_c|\vec{k}|,\ y=\rho_c^2M_X^2$. This
cross section can be compared with the leading order pQCD cross
section
\be \frac{zd\sigma^{qq \rightarrow
qq\pi^0}}{dzd^2\vec{k}}(pQCD) =\frac{D_q^{\pi^0}(z)}{\pi
z}\frac{d\sigma}{dq^2},
\label{pQCD}
\ee where $D_q^{\pi^0}(z)$ is
the fragmentation function and
$d\sigma/dq^2=8\pi\alpha_s^2(q^2)/(q^2-m_g^2)^2/9$.
We should mention that in the kinematic region of few GeV for
the transverse momentum of the pion and at high energy
one can neglect the change of the longitudinal momentum
of the quark in partonic subprocess in the pQCD. In this case,
the meaning of $z$ becomes the same for both pQCD and APP cases.
In Fig.3, the result of calculation of the cross section of the
APP, Eq.\ref{partonicAPP}, in the quark-quark scattering is
presented \footnote{The pion production by our nonperturbative
mechanism in the quark-gluon scattering is scaled by the same color
factor $9/4$ as in the pQCD case. Therefore, the inclusion into
consideration of the quark-gluon scattering does not influence
 our conclusion. We also don't consider the pQCD contribution
coming from the $gg\rightarrow gg$ subprocess, because it gives the
sizable contribution only at very small value of $z$. The pQCD
contribution from the  gluonic subprocess $gg\rightarrow gg$ is
expected to be small in the  large $x_F$ region even for the RHIC
energies \cite{kretzer}.}. We compare it with the pQCD cross
section, Eq.\ref{pQCD}, calculated with the strong coupling constant
given by the analytical pQCD \cite{Shirkov:1997wi}
\be
\alpha_s(Q^2)=\frac{1}{\beta_0}\bigg[\frac{1}{log(Q^2/\Lambda^2)}+\frac{\Lambda^2}{\Lambda^2-Q^2}\bigg],
\label{coupling} \ee
where $Q^2=-q^2$, $\beta_0=(33-2N_f)/12\pi$, $\Lambda=250$ MeV,
and $N_f=3$ were used. The fragmentation
function $D_q^{\pi^0}(z)$ was taken from \cite{Kniehl:2000fe}, set
KKP-1. The dynamical gluon mass  $m_g\approx 0.65$
GeV was fixed according to the result of
the lattice data in  \cite{RuizArriola:2004en}\footnote{We do not consider the
possible effect coming from the running dynamical gluon mass.}.

\begin{widetext}

\begin{figure}[h]
\begin{minipage}[c]{8cm}
%\vskip -0.5cm
\hspace*{-3.0cm}
\centering{\includegraphics[width=5cm,height=5cm,angle=0]{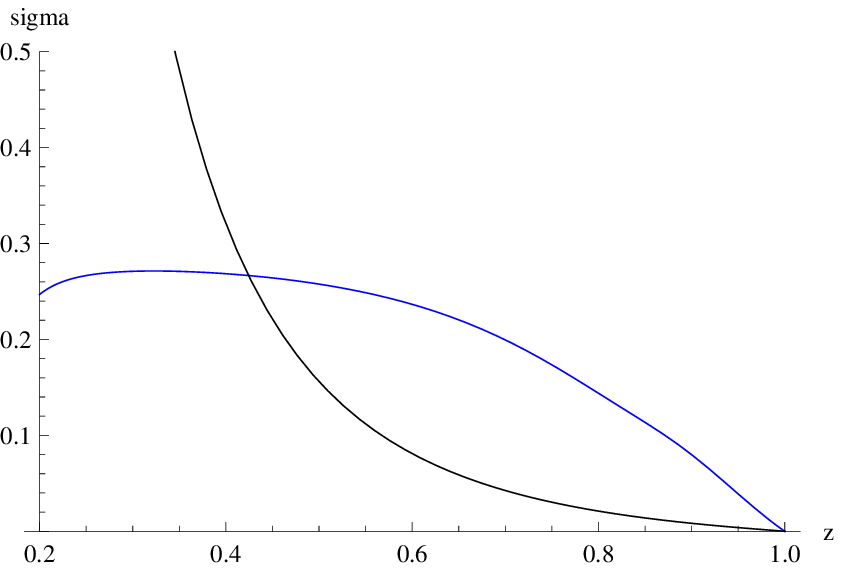}}\
\end{minipage}
\begin{minipage}[c]{8cm}
\hspace*{-7.0cm}
 %\vskip -1cm
\centering{\includegraphics[width=5cm,height=5cm,angle=0]{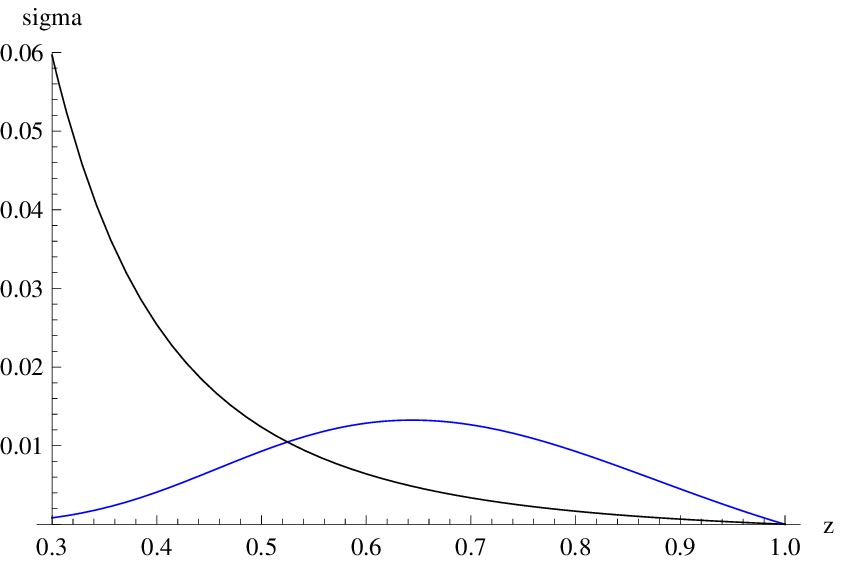}}\
%\hspace*{1.0cm}
 %\vskip -1cm
\end{minipage}
\begin{minipage}[c]{8cm}
\vskip -5cm \hspace*{8.0cm}
\centering{\includegraphics[width=5cm,height=5cm,angle=0]{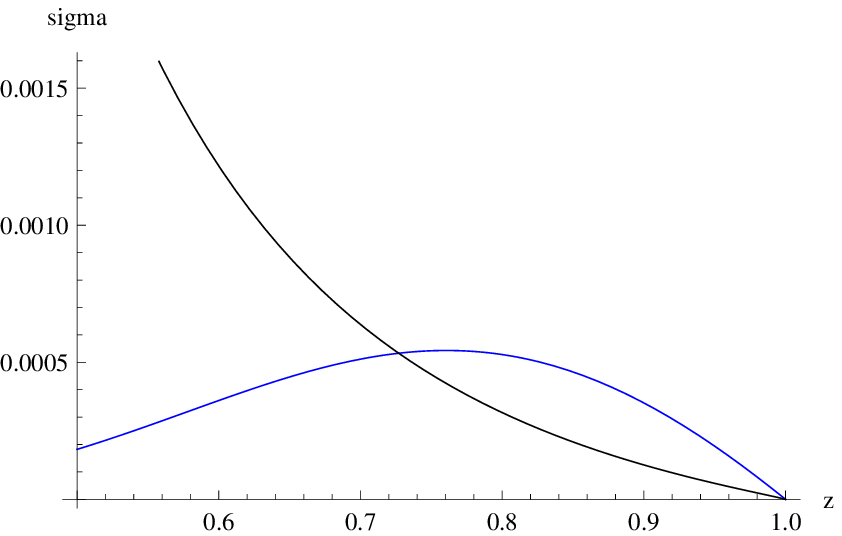}}\
%\hspace*{1.0cm}
\end{minipage}
\caption{ The $z$ dependency of the pQCD (black line) and the  APP (blue line) cross sections,
in the units $10^2\mu b/GeV^2$, for the different values of the pion transverse momentum:
the left panel $k_t=1$ GeV, the central panel $k_t=2$ GeV,
and the right panel $k_t=3$ GeV. Here, the relation $k_t=|\vec{k}|\approx \sqrt{-q^2}$ was used.}
\end{figure}
\end{widetext}
Let us discuss, in the beginning on a qualitative level, the features of
kinematics of the pQCD and the APP contributions to $\pi^0$ production in the
fixed target experiments for a few GeV transverse momentum of
the final pion where the big difference from the pQCD prediction was
observed. In this case, the main contribution comes from the
scattering of the valence quarks which carry about $<x>\approx
0.2$ momentum of the initial proton. Therefore, it follows from
Fig.3 that at $k_t=1$ GeV the APP contribution starts to be
dominant at $x_F\approx 0.08$ \footnote{ The $x_F$ is  part of
the initial proton momentum carried by the final pion in the center of
mass system in the longitudinal direction.}, at $k_t=2$ GeV
the APP contribution dominates at $x_F\approx 0.11$, and at
$k_t=3$ GeV the APP above $x_F\approx 0.15$ gives the
main contribution to the cross section. So we come to the
conclusion that the pQCD might give the main contribution only in the
region of very small $x_F$. Furthermore, the APP contribution is
very big even at large $k_t$ in the large $x_F$ region. To
calculate the APP cross section at the hadron level, we use an
approach similar to the pQCD approach (see \cite{field}). In
particular, in the high energy limit we will neglect all masses of
the hadrons and partons in the kinematics of the inclusive pion production.
In this way, for the APP contribution to the inclusive cross
section of the $\pi^0$ production in the proton-proton scattering, we have

\begin{widetext}
\be
\frac{E_kd\sigma}{d^3k}=\sum_{a,b}\int_{x_a^{min}}^1dx_a\int_{x_b^{min}}^1G_{A\rightarrow
a}(x_a,\mu^2) G_{B\rightarrow
b}(x_b,\mu^2)\frac{zd\sigma^{ab}}{dzd^2\vec{k}}(z=z_c,k_t),
 \label{APP}
\ee
\end{widetext}
where
$G_{A,B\rightarrow a,b}$ are parton distribution functions (PDFs) and the sum over the different types of partons is carried out.
Additionally, we should use the scale $\mu^2\approx 1/\rho_c^2$ in our nonperturbative
calculation.
\begin{figure}[h]
\begin{minipage}[c]{8cm}
%\vskip -0.5cm
\hspace*{-1.0cm}
\centering{\includegraphics[width=5cm,height=5cm,angle=0]{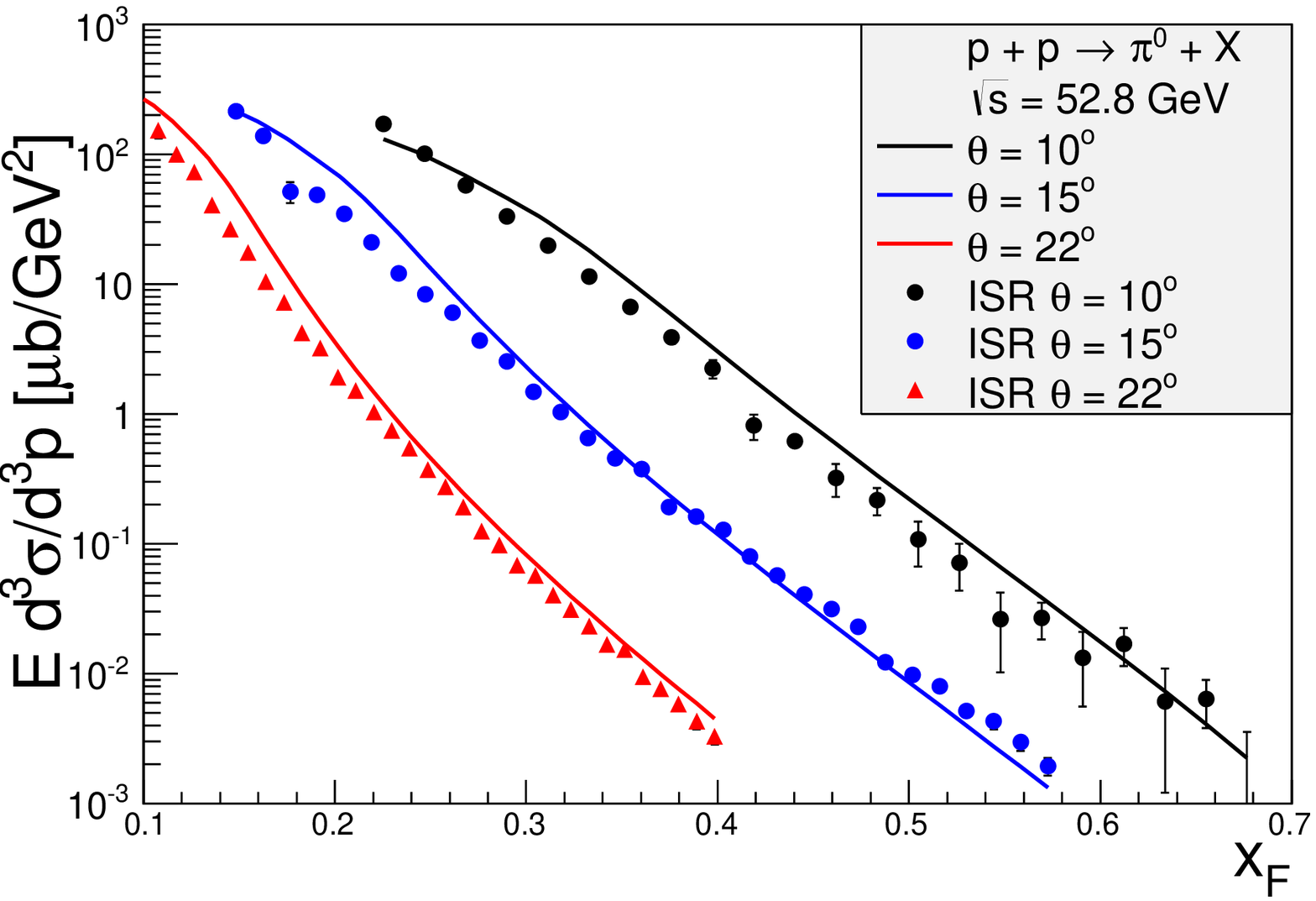}}\
\end{minipage}
\begin{minipage}[c]{8cm}
\hspace*{-1.0cm}
 %\vskip -1cm
\centering{\includegraphics[width=5cm,height=5cm,angle=0]{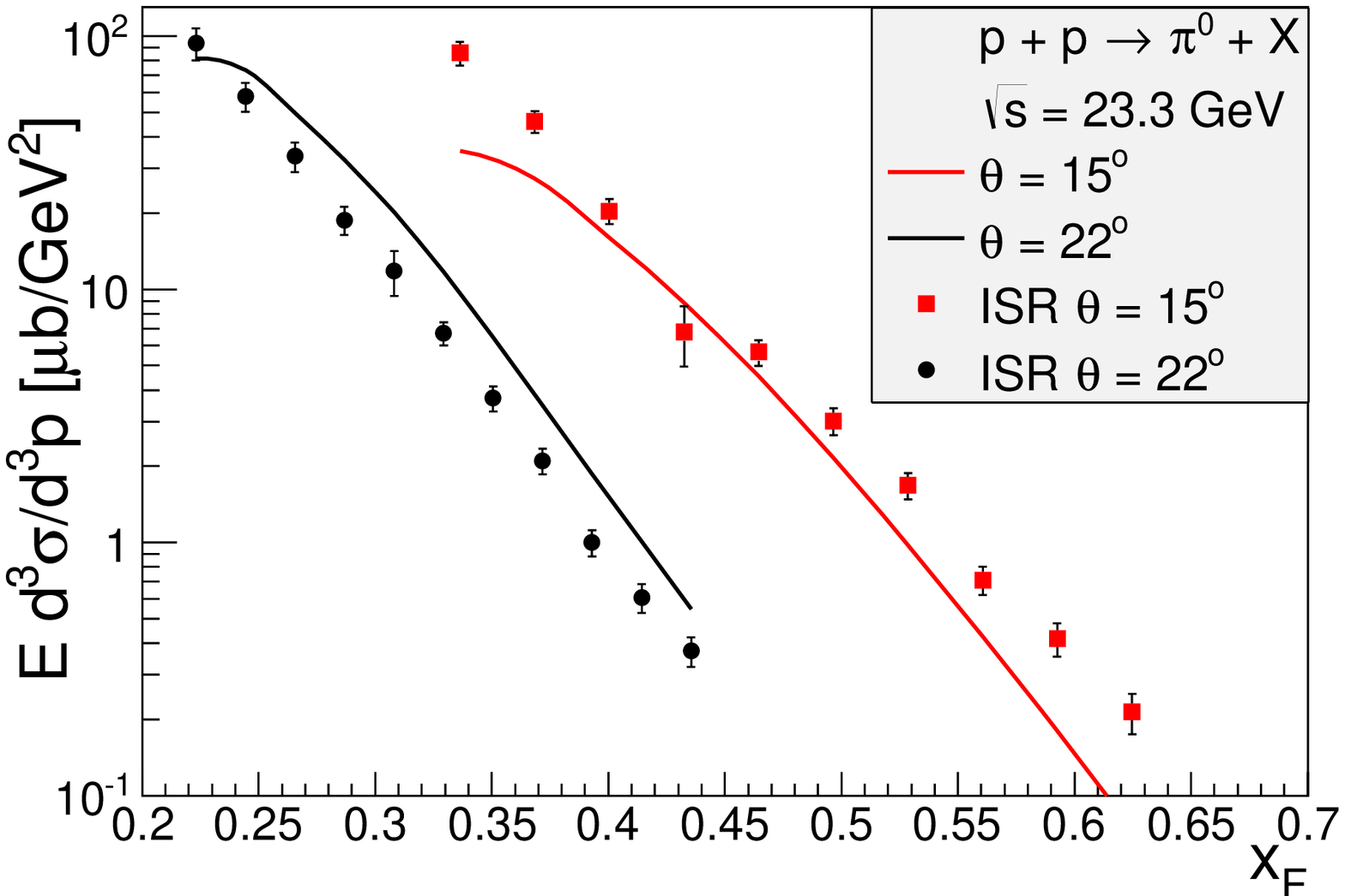}}\
%\hspace*{1.0cm}
 %\vskip -1cm
\end{minipage}
\begin{minipage}[c]{8cm}
\vskip 0cm
\hspace*{-1.0cm}
\centering{\includegraphics[width=5cm,height=5cm,angle=0]{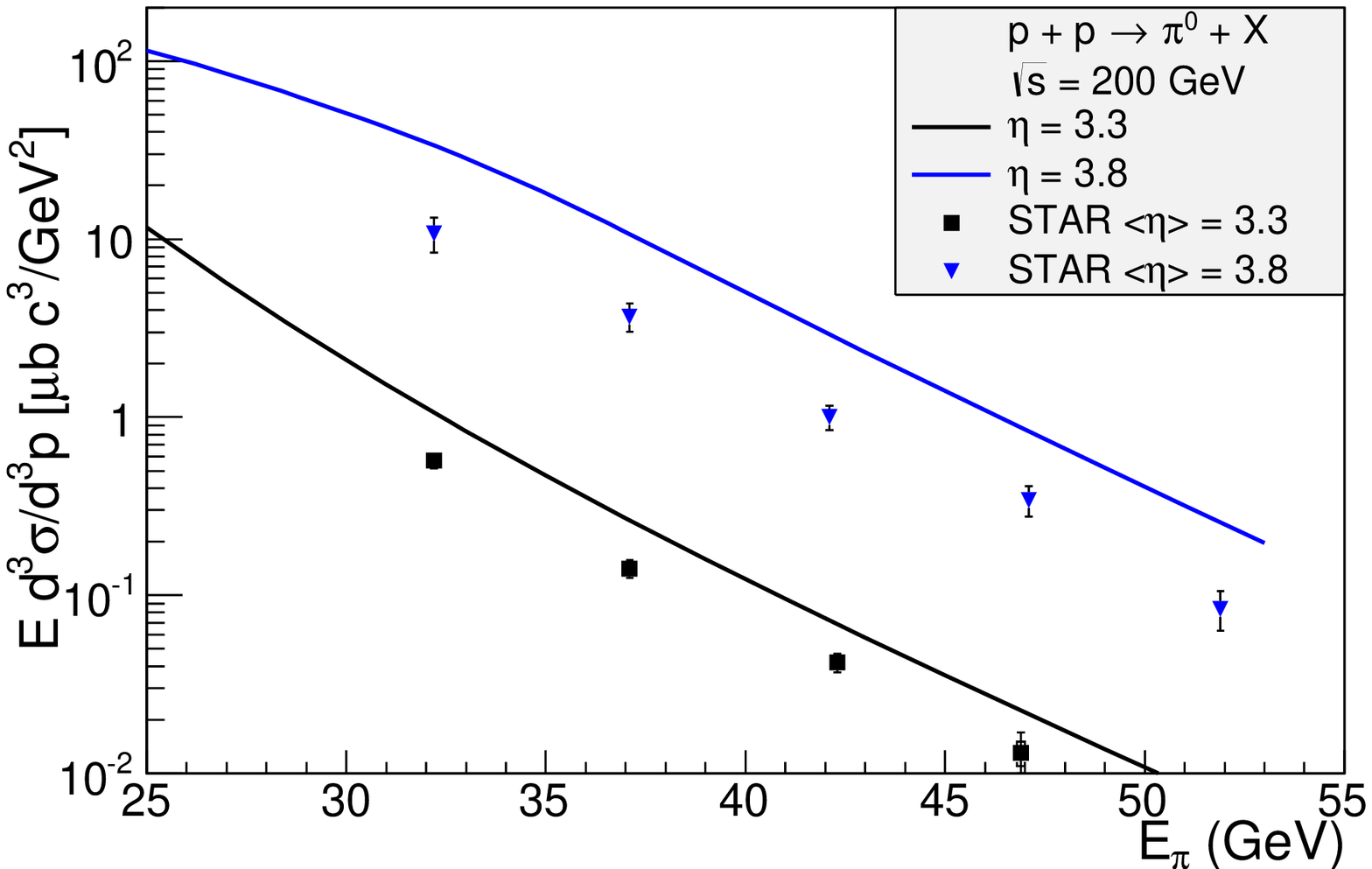}}\
%\hspace*{1.0cm}
\end{minipage}
\caption{The APP cross sections for inclusive $\pi^0$ in
comparison with the ISR (CERN Intersecting Storage Rings) fixed target \cite{ISR} and the STAR
\cite{STAR} data for $k_t\approx 1\div 3$ GeV at large pseudorapidity.}
\end{figure}

In Eq.\ref{APP}
\ba
x_a^{min}&=&x_1/(1-x_2), \  \ x_b^{min}=x_2x_a/(x_a-x_1),\nonumber \\
x_1&=&\frac{1}{2}x_T/tan(\theta_{cm}/2),\  \  x_2=\frac{1}{2}x_Ttan(\theta_{cm}/2), \nonumber\\
z_c&=&x_2/x_b+x_1/x_a, \label{kin}
\ea
where $x_T=2k_t/\sqrt{s}$
and $\theta_{cm}$ is the c.m. angle of the produced pion and
the GRV98 NLO parametrization at $\mu^2=0.4$ GeV$^2$
was used \cite{Gluck:1998xa}. This value of the scale is very
close to the scale of our instanton based calculation $1/\rho_c^2=
0.36$ GeV$^2$. The results are represented in
Fig.4 in  comparison with the experimental data. As one can see, the APP
mechanism provides a rather good description of both fixed target
and collider data at $x_F>0.2$ and few GeV value for the transverse
momentum of the pion. Some deviation from STAR data might be related
to the PDF uncertainties. On the other hand, it is known that in
this kinematic region for the fixed target experiments the pQCD
contribution is very small and cannot explain the data without
introducing large intrinsic parton momentum \cite{soffer,alesio}
in the leading order calculation. The NLO pQCD predictions also fail
to explain these data. We would like to emphasize that it is very
difficult to uncerstand in the conventional approach why the pQCD works
well at RHIC energy but fails to explain large $x_F$  ISR data.
In both the cases the energy is much larger than
any other scales, i.e., hadron masses, transverse momentum of
partons and hadrons, transfer momentum in a partonic subprocess, etc.
Furthermore, the contribution from gluonic $gg\rightarrow gg$,
which could be one of the candidates to explain this difference,
is expected to be small in the forward rapidity region even for the
RHIC energies \cite{kretzer}. It might be that a rather good
agreement for the collider case \cite{Boglione:2007dm} is related to
the internal uncertainties in the pQCD calculation due to
uncertainties in the value of factorization scale and in the
shape of fragmentation functions. Indeed, the comparison with the
STAR data shows rather large sensitivity of the NLO pQCD prediction
to the choice of the fragmentation function \cite{STAR} and to the
factorization scale \cite{BRAHMS}.
 Therefore, we believe that the complete picture of the inclusive
particle production in the strong interaction should include both
pQCD and APP mechanisms.

\section{Conclusion}
We discussed the new mechanism for the inclusive production of the
pions in the hadron-hadron interaction. It is related to the
anomalous quark-gluon-pion coupling induced by the nontrivial
topological structure of the QCD vacuum. It is shown that this APP
mechanism gives the dominant contribution to the inclusive pion
production for the $x_F\gtrsim 0.1$ in the few GeV region for
the transverse momentum of the pion. Furthermore, the APP violates
factorization in the inclusive pion production and allows one to calculate
the cross section without any fragmentation functions.
The APP mechanism could be extended to the other pseudoscalar
mesons. In the case of $K$- and $\eta$-meson production, it is
needed to introduce the suppression for the chromomagnetic interaction
due to the large mass of the $s$-quark. Additionally, one should
include the effect from the $\eta-\eta^\prime$ mixing in the
consideration of $\eta$-meson production. It is evident that the
APP mechanism should give the contribution to other observables in
the pseudoscalar meson production in high energy reactions with hadrons.
We should emphasize that the anomalous quark-gluon-pion coupling,
Eq.\ref{Lag2}, flips the quark spin.
Therefore, it should contribute to the large single spin
asymmetries (SSA) in the inclusive pseudoscalar meson production
observed in the different high energy reactions with hadrons. The
other longstanding problem is the large transverse polarization of $\Lambda$
polarization  observed in different high energy reactions (see
discussion in \cite{Anselmino:2000vs,Anselmino:2002rr}). In our
approach, such polarization might come from quark spin-flip in
the $K$-meson production by the APP-type mechanism, e.g. $u+q(g)\rightarrow
K^++s+q(g)$, and following fragmentation of the strange quark to
$\Lambda$. In this case, the SSA in the $K$-meson production and the transverse
polarization of $\Lambda$ should be deeply related to each other.
However, the calculation of SSA is a much more complicated task due
to the loop diagrams contribution even in the leading order.
The APP mechanism is completely different from the
mechanism related to the first term in Eq.\ref{Lag3} which was considered  in
\cite{Kochelev:2006ny,kochelev3} at the partonic level. However, at
the hadron level one should include the fragmentation function
into consideration. It is clear that such a contribution to
the pseudoscalar meson inclusive cross section should have
approximately the same $x_F$ dependency as the pQCD contribution
which cannot solve the problem of the pion production at large
$x_F$. Our separate task is to study the effects of the
APP mechanism to the inclusive production of the pseudoscalar
mesons in high energy heavy ion collision, in particular, to the
so-called nuclear modification factor $R_{AA}$.
In the case of pQCD production, the size of the pion is
fixed by the fragmentation dynamics and should be rather large,
$R_\pi^{pQCD}\approx 1 $ fm. For the APP mechanism the size of the pion is
small, $R_\pi^{APP}\approx \rho_c\approx 0.3$ fm.  It follows that
the final state interaction cross section of the pion in the nuclear
matter for the APP process is about one order of  magnitude
smaller in comparison with the pQCD case. Therefore, the pQCD and the
APP mechanisms should have different nuclear dependency.
This difference, for example, might be responsible for the peak in
$R_{AA}$ observed at RHIC and LHC in the few GeV region for transverse
momentum of the final mesons.
The work in these directions is now in  progress.

\section{Acknowledgments}
We would like to thank A.E. Dorokhov and S.B. Gerasimov   for
useful discussions. We are very grateful to Umberto D'Alesio for
providing us with the tables of  experimental data for the cross
section of the inclusive $\pi^0$ production in different high
energy reactions. This work was partially supported by the
National Natural Science Foundation of China (Grant No. 11035006
and 11175215), and by the Chinese Academy of Sciences visiting
professorship for senior international scientists (Grant No.
2013T2J0011). This research was also supported in part by the
Basic Science Research Program through the National Research
Foundation of Korea(NRF) funded by the Ministry of
Education(2013R1A1A2009695)(HJL).

\end{document}